\documentclass{appolb}
\usepackage{epsfig}
\usepackage{amsmath}
\newcommand{\beq}{\begin{equation}}
\newcommand{\eeq}{\end{equation}}
\newcommand{\beqa}{\begin{eqnarray}}
\newcommand{\eeqa}{\end{eqnarray}}
\newcommand{\bseq}{\begin{subequations}}
\newcommand{\eseq}{\end{subequations}}

\def\x{{\boldsymbol x}}

\def\p{{\boldsymbol p}}
\def\0{{\boldsymbol 0}}

\def\simle{\mathrel{\rlap{\raise 0.511ex \hbox{$<$}}{\lower 0.511ex 
\hbox{$\sim$}}}}
\def\simge{\mathrel{ \rlap{\raise 0.511ex 
\hbox{$>$}}{\lower 0.511ex \hbox{$\sim$}}}}

\begin{document}
\pagestyle{plain}
\eqsec
\newcount\eLiNe\eLiNe=\inputlineno\advance\eLiNe by -1
\title{Heavy quarks in nucleus-nucleus collisions:\\
from RHIC to LHC}
\author{A. Beraudo$^{1,2}$, W.M. Alberico$^{3,4}$, A. De Pace$^4$, A. Molinari$^{3,4}$, M. Monteno$^4$, M. Nardi$^4$ and F. Prino$^4$
\address{$^1$Centro Studi e Ricerche \emph{Enrico Fermi}, Piazza del Viminale 1, Roma, Italy}
\address{$^2$Physics Department, Theory Unit, CERN, CH-1211 Gen\`eve 23, Switzerland}
\address{$^3$Dipartimento di Fisica Teorica dell'Universit\`a di Torino and\\
$^4$Istituto Nazionale di Fisica Nucleare, Sezione di Torino,\\
via P.Giuria 1, I-10125 Torino, Italy} 
}
\maketitle

\begin{abstract}
We present a study of the heavy-flavor dynamics in nucleus-nucleus collisions. The initial (hard) production of $c$ and $b$ quarks is taken from NLO pQCD predictions. The presence of a hot medium (a Quark Gluon Plasma described by hydrodynamics) affects the final spectra of open-charm (beauty) hadrons and their decay electrons with respect to what found in $pp$ collisions. The propagation of $c$ and $b$ quarks in the plasma is based on a picture of multiple uncorrelated random collisions, described by a relativistic Langevin equation. A microscopic evaluation of the transport coefficients is provided within a pQCD approach (with proper resummation of medium effects). Results for the final spectra of heavy-flavor hadrons and decay-electrons are given, with particular emphasis on $R_{AA}$ and $v_2$.
\end{abstract}

\section{Introduction}
Heavy flavor spectra are an interesting probe of the medium formed in heavy-ion collisions: $c$ and $b$ quarks are produced in hard processes during the crossing of the two nuclei and the QCD flavor conservation allows to ``tag'' final-state charm and beauty hadrons, so that the color charge and mass of the parent parton propagating in the plasma is known.
The interaction between the heavy quarks and the medium leads to a modification of the final spectra with respect to the $pp$ case. Different mechanisms and approaches were proposed in the literature to account for such an effect: \eg medium-induced gluon-radiation~\cite{gyu,asw}, scattering mediated by resonant states~\cite{rapp}, calculations based on the Boltzmann equation~\cite{bamps}.
As explained at length in~\cite{langepap,langepro} we describe the heavy-quark dynamics through the relativistic Langevin equation. This amounts to assume that the heavy-quark momentum, crossing the plasma, gets changed due to multiple uncorrelated random collisions. For independent studies based on the Langevin equation see for instance Refs.~\cite{tea,aic,hira}.
\section{The relativistic Langevin equation}
\begin{figure}
\includegraphics[clip,width=0.48\textwidth]{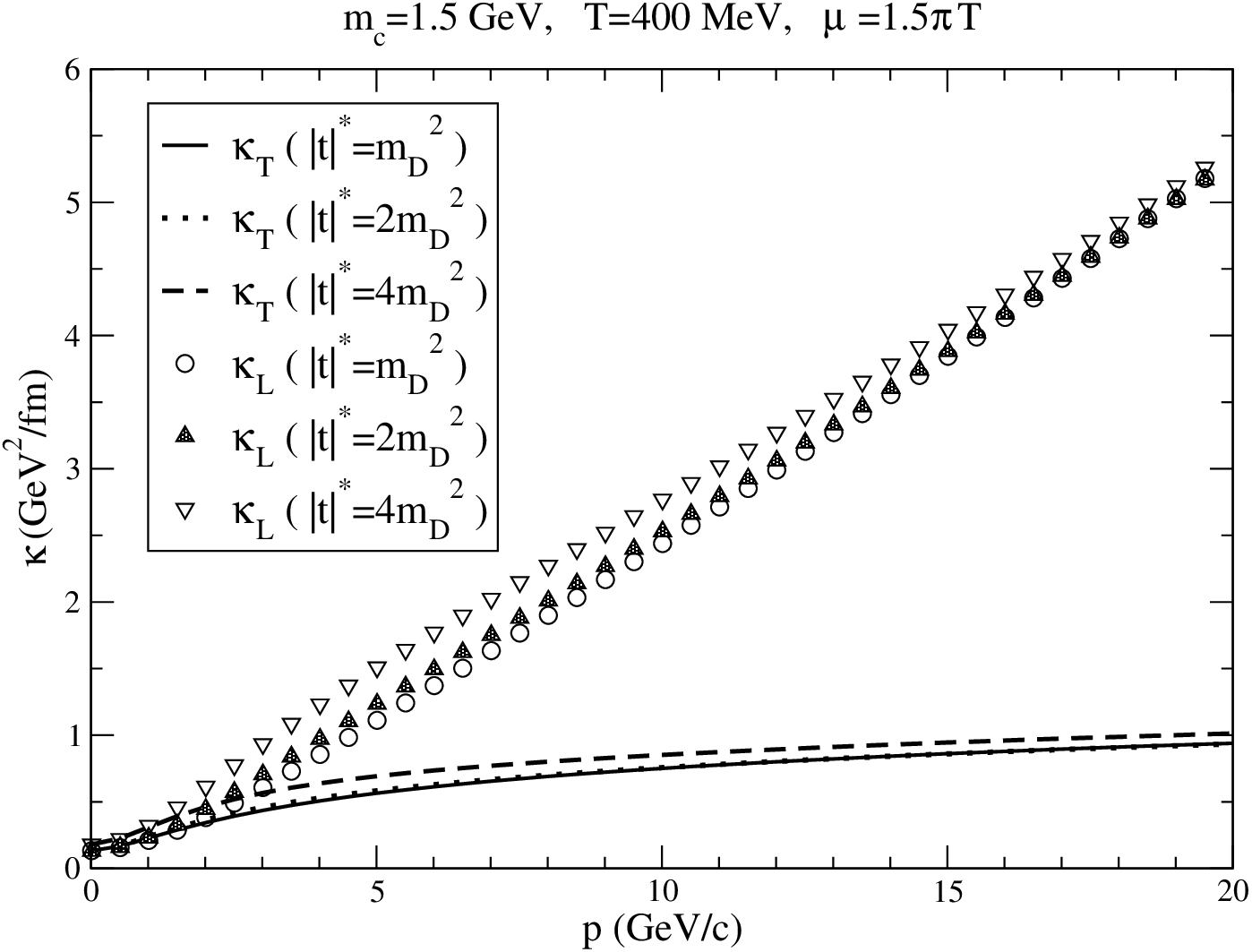}
\includegraphics[clip,width=0.48\textwidth]{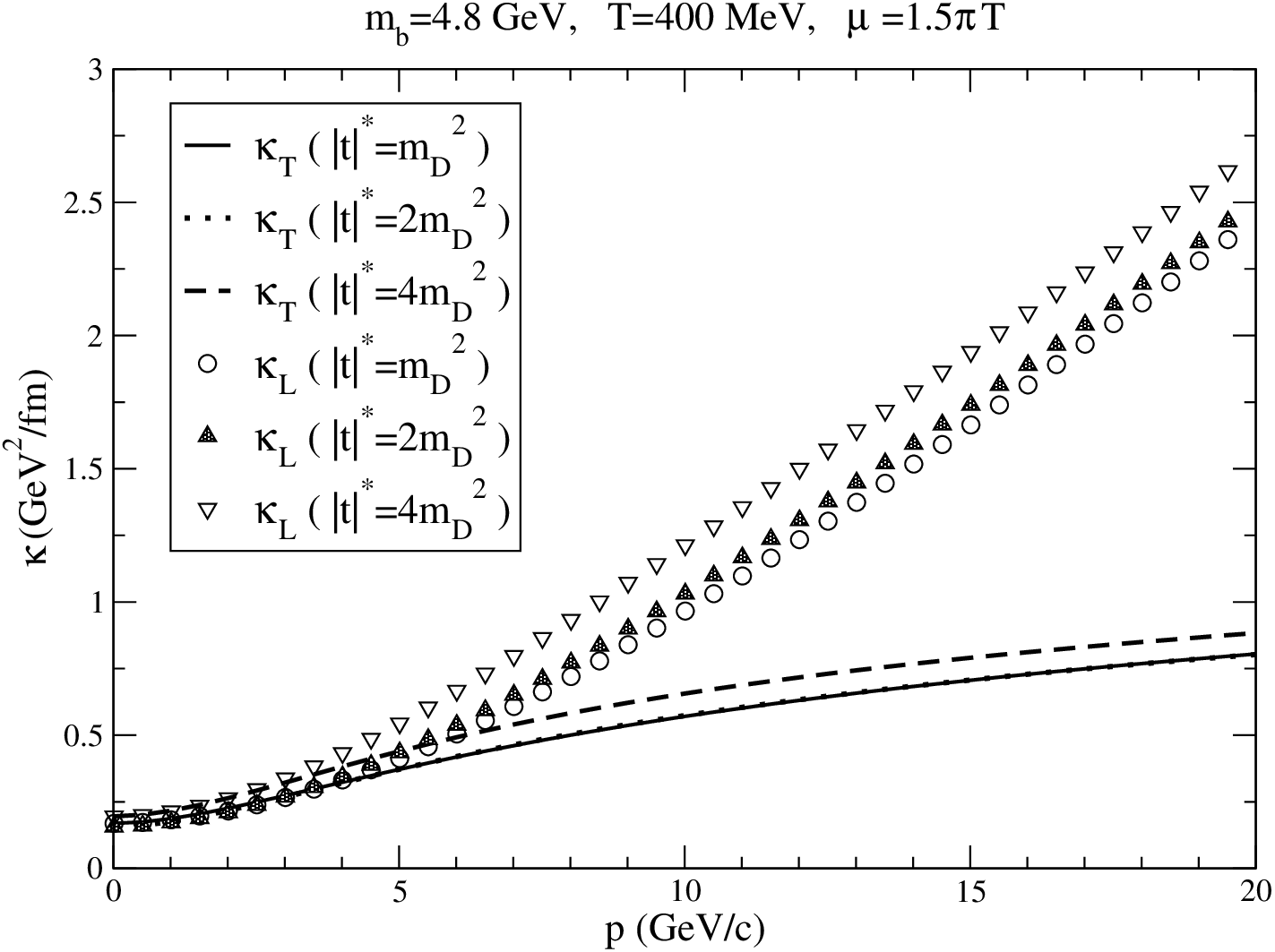}
\caption{The $c$ (left panel) and $b$ (right panel) momentum diffusion coefficients $\kappa_{T/L}(p)$ resulting from the sum of the soft and hard contributions. The sensitivity to the intermediate cutoff $|t|^*\sim m_D^2$ is very mild. The curves refer to the temperature $T=400$~MeV, with the coupling $g$ evaluated at the scale $\mu=(3/2)\pi T$.}
\label{fig:transpcoeff}
\end{figure}
The relativistic Langevin equation~\cite{lange0} is used to study the propagation of $c$ and $b$ quarks in the QGP. The variation of the heavy-quark momentum in the time-interval $\Delta t$
\beq\label{eq:lange_r_d}
\frac{\Delta p^i}{\Delta t}=-\eta_D(p)p^i+\xi^i(t),
\eeq
is given by the sum of a \emph{deterministic} friction term and a \emph{stochastic} noise term $\xi^i(t)$, which is completely determined by its two-point temporal correlator  
\beq\label{eq:noise1}
\langle\xi^i(t)\xi^j(t')\rangle=b^{ij}(\p)\delta(t-t'),\quad{\rm with}\quad
b^{ij}(\p)\equiv \kappa_L(p)\hat{p}^i\hat{p}^j+\kappa_T(p)
(\delta^{ij}-\hat{p}^i\hat{p}^j).
\eeq
The latter involves the transport coefficients $\kappa_T(p)\!\equiv\!\frac{1}{2}\frac{\langle \Delta p_T^2\rangle}{\Delta t}$ and $\kappa_L(p)\!\equiv\!\frac{\langle \Delta p_L^2\rangle}{\Delta t}$,
representing the average transverse and longitudinal squared-momentum acquired per unit time by the quark due to the collisions in the medium.

The transport coefficients $\kappa_{T/L}$ are evaluated according to the procedure presented in~\cite{langepap}. We introduce an intermediate cutoff $|t|^*\!\sim\!m_D^2$ ($t\!\equiv\!(P'\!-\!P)^2$) separating hard and soft scatterings. The contribution of hard collisions ($|t|\!>\!|t|^*$) is evaluated through a kinetic pQCD calculation of the processes $Q(P)q_{i/\bar i}\!\to\! Q(P')q_{i/\bar i}$ and $Q(P)g\!\to\! Q(P')g$. On the other hand in soft collisions ($|t|\!<\!|t|^*$) the exchanged gluon feels the presence of the plasma. A resummation of medium effects is thus required and this is provided by the Hard Thermal Loop approximation. The final result is given by the sum of the two contributions $\kappa_{T/L}(p)=\kappa_{T/L}^{\rm hard}(p)+\kappa_{T/L}^{\rm soft}(p)$ and its explicit expression can be found in Ref.~\cite{langepap}. In Fig.~\ref{fig:transpcoeff} we display the behavior of $\kappa_{T/L}$ for $c$ and $b$ quarks. The sensitivity to the intermediate cutoff $|t|^*$ is quite small, supporting the validity of the approach. 
\section{Heavy flavor in $pp$ and heavy-ion collisions}
The procedure to generate final-state heavy-flavor spectra can be factorized into some independent steps.
Details can be found in~\cite{langepap}. 
\begin{itemize}
\item \emph{Initial production of $Q\overline{Q}$ pairs in hard pQCD processes}.
An initial sample of $c$ and $b$ quarks is generated using the POWHEG code~\cite{POWHEG}, which implements pQCD at NLO. In the $AA$ case nuclear PDFs are employed~\cite{EPS09}. Heavy quarks are then distributed in the transverse plane according to the nuclear overlap function ${dN/d\x_\perp\!\sim\!T_{AB}(x,y)}\!\equiv\!T_A(x\!+\!b/2,y)T_B(x\!-\!b/2,y)$ and given a $k_T$-broadening depending on the crossed thickness of nuclear matter.
\item \emph{Langevin evolution in the fireball}. In the $AA$ case at the proper-time $\tau\!\equiv\sqrt{t^2\!-\!z^2}\!=\!\tau_0$ one starts following the stochastic Langevin dynamics of the quarks in the plasma. The expansion of the background medium is described by ideal/viscous hydrodynamics~\cite{kolb1,rom1,rom2}.
\item \emph{Hadronization and decays}. Heavy quarks are made hadronize (around the critical energy density $\epsilon_c$ in the AA case) using Peterson fragmentation functions~\cite{peter} with branching fractions into hadrons
taken from Refs.~\cite{zeus,pdg}. Finally each hadron is forced to decay into electrons with PYTHIA~\cite{Pythia}, using updated tables of branching ratios~\cite{pdg10}. 
\end{itemize}
\section{Numerical results}
\begin{figure}
\includegraphics[clip,width=0.48\textwidth]{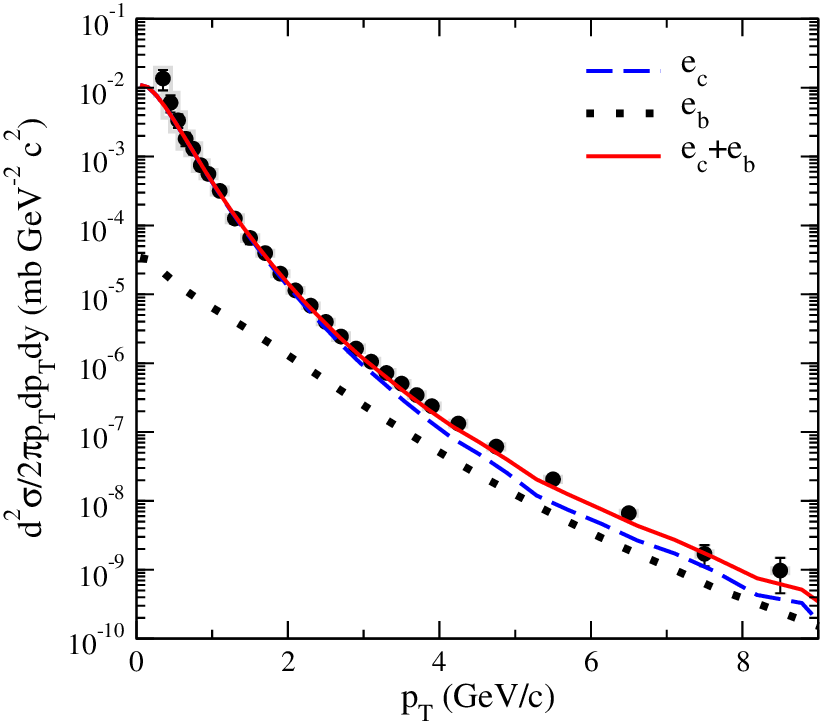}
\includegraphics[clip,width=0.48\textwidth]{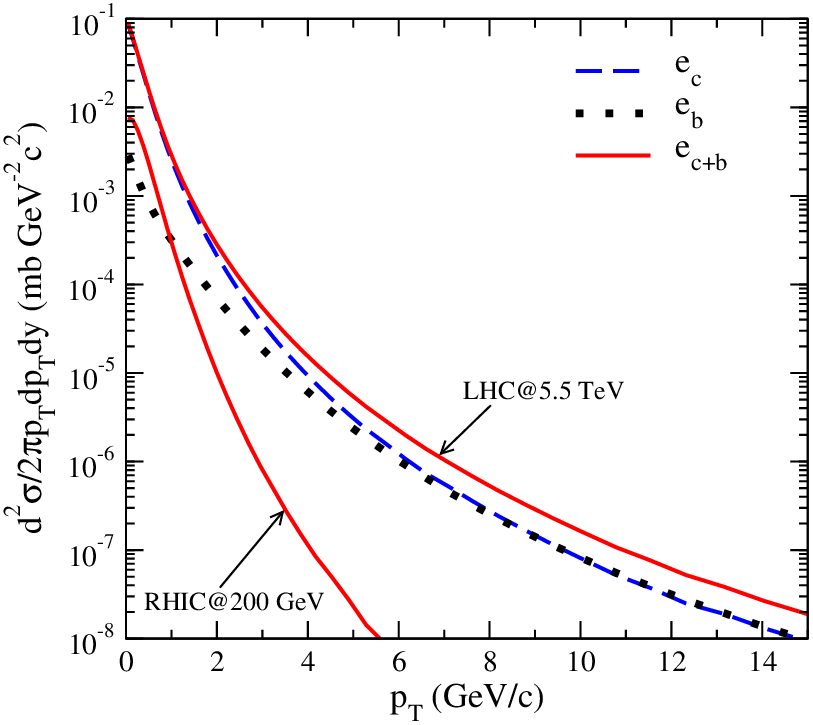}
\caption{Invariant differential cross section of electrons [$(e^++e^-)/2$],
from heavy-flavor decay in $pp$ collisions at $\sqrt{s}=200$~GeV (left panel) and $\sqrt{s}=5.5$~TeV (right panel). The curves refer to electrons from $c$ or $b$ quarks generated by POWHEG. PHENIX data are extremely well reproduced.}
\label{fig:ppspectrum}
\end{figure}
Our predictions for the cross sections of heavy-flavor electrons in $pp$ collisions at RHIC ($\sqrt{s}\!=\!200$ GeV) and LHC ($\sqrt{s}\!=\!5.5$ TeV) are given in Fig.~\ref{fig:ppspectrum}. PHENIX results are well reproduced. Going from RHIC to LHC one observes a sizable increase of the cross section, a hardening of the spectrum and the growth of the relative contribution of electrons from $b$ decays.

\begin{figure}
\begin{center}
\includegraphics[clip,width=0.48\textwidth]{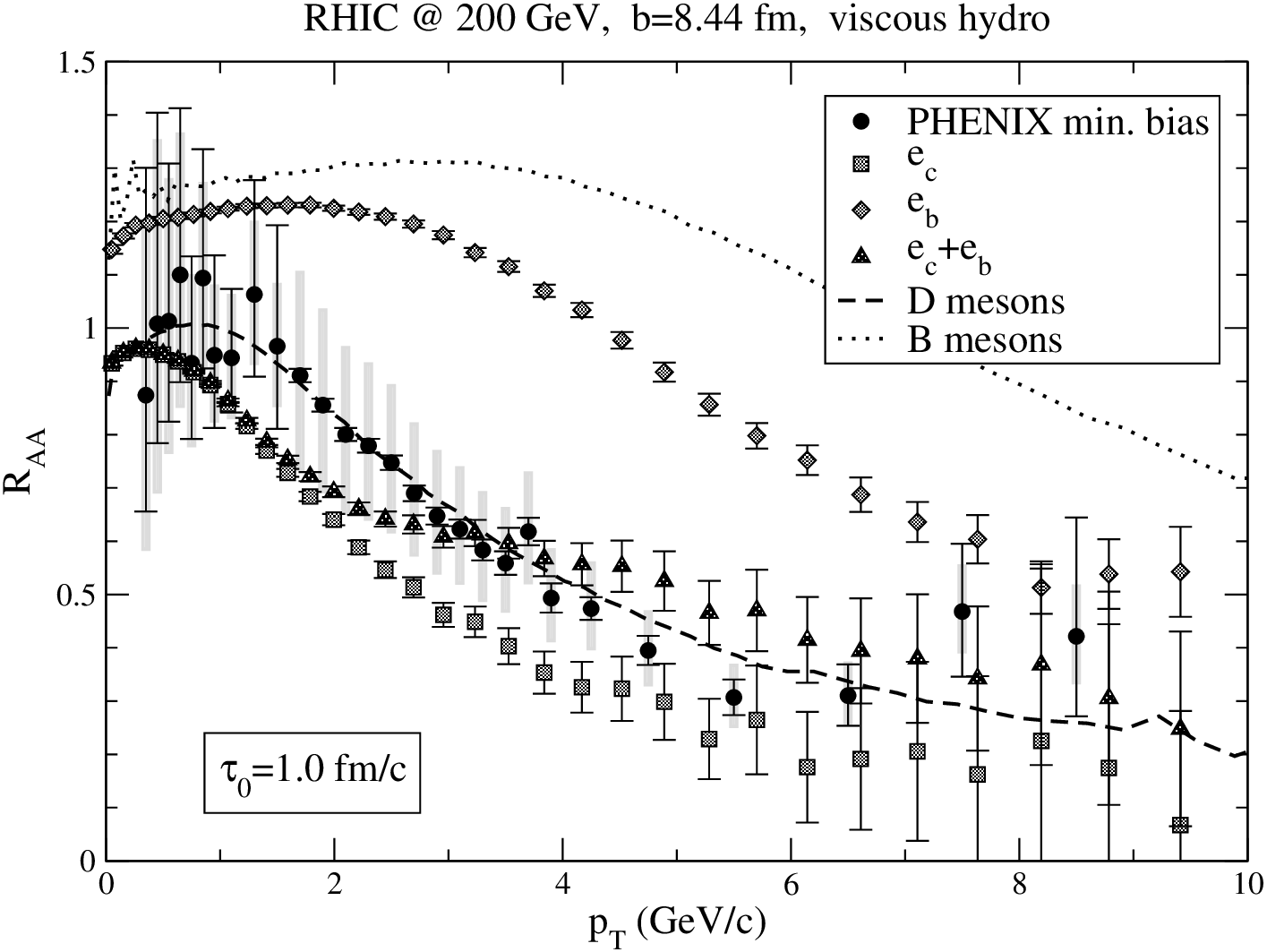}
\includegraphics[clip,width=0.48\textwidth]{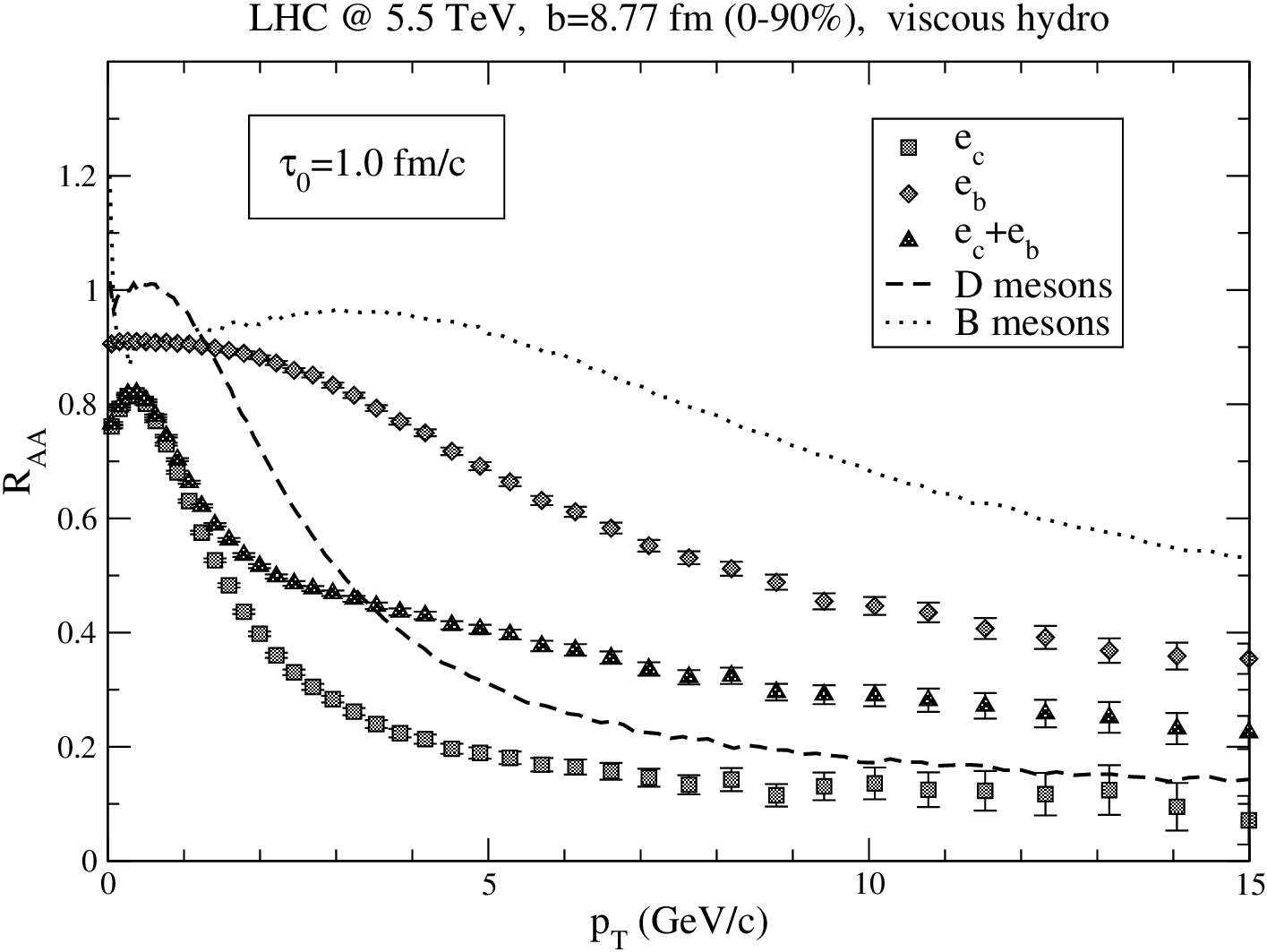}
\caption{The $R_{AA}$ of heavy-flavor hadrons and decay-electrons in minimum-bias AA collisions at RHIC ($0-92\%$ of the total hadronic cross section, left panel) and LHC@5.5~TeV ($0-90\%$, right panel) for viscous hydrodynamics.}
\label{fig:RAA_pT}
\end{center}
\end{figure}
\begin{figure}
\begin{center}
\includegraphics[clip,width=0.9\textwidth]{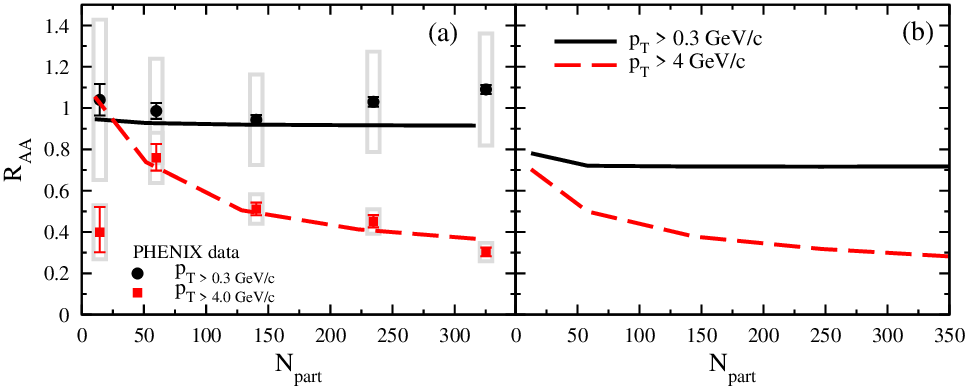}
\caption{The nuclear modification factor of heavy-flavor electrons at RHIC (a) and LHC@5.5~TeV (b) for viscous hydrodynamics ($\tau_0=1$~fm), obtained by integrating the electron yields over the indicated momentum ranges, as a function of $N_{\rm part}$. The centrality dependence found by the PHENIX experiment \cite{phenix2} is nicely reproduced.}
\label{fig:RAA_int}
\end{center}
\end{figure}
For $AA$ collisions effects of the Langevin evolution where studied through the nuclear modification factor $R_{AA}(p_T)\!\equiv\!(dN/dp_T)^{AA}/\langle N\rangle_{\rm coll} (dN/dp_T)^{pp}$ and the elliptic flow coefficient $v_2(p_T)\!\equiv\!\langle\cos(2\phi)\rangle_{p_T}$. Here we display results obtained using viscous hydrodynamics, for few representative cases among the ones explored in~\cite{langepap}.
In Fig.~\ref{fig:RAA_pT} we show our findings for the $R_{AA}(p_T)$ of heavy-flavor hadrons and their decay-electrons at RHIC and LHC@5.5~TeV. A general agreement with PHENIX data \cite{phenix2} can be attained.
Notice that the higher temperatures reached at LHC lead to a larger quenching of the spectra; however in the inclusive ($e_c\!+\!e_b$) result the effect is partially compensated by the stronger weight acquired by the $b$ contribution.
In Fig.~\ref{fig:RAA_int} the centrality dependence of  $R_{AA}$ is displayed. The overall agreement with PHENIX data looks satisfactory.     

\begin{figure}
\begin{center}
\includegraphics[clip,width=0.85\textwidth]{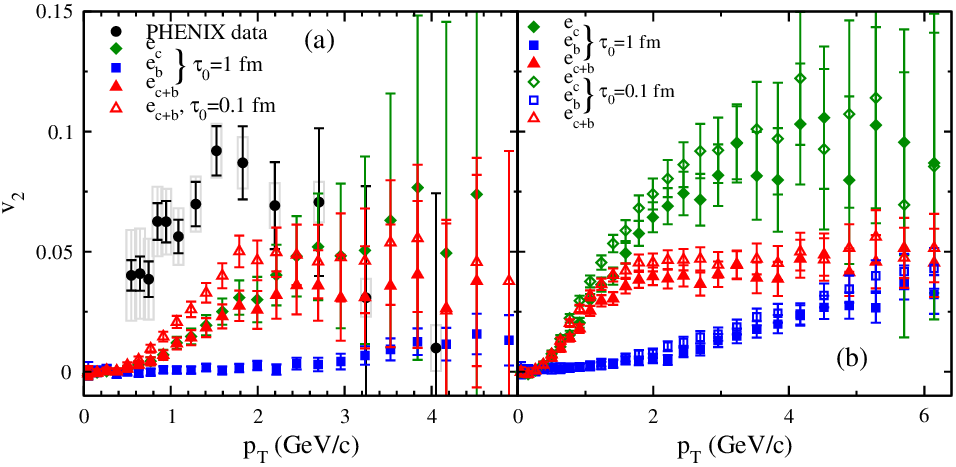}
\caption{(a) The minimum-bias anisotropy parameter $v_2$ for heavy-flavor electrons at RHIC for viscous hydrodynamics. PHENIX data\protect\cite{phenix,phenix2} are compared with the outcomes of our calculations for electrons from $c$ and $b$ quarks and their combination. (b) As in panel (a), but for LHC and two choices of $\tau_0$.}
\label{fig:v2}
\end{center}
\end{figure}
In Fig.~\ref{fig:v2} we display our results for $v_2$. Even assuming a very rapid thermalization, our findings tend to underestimate the experimental data: perturbative cross sections favor small-angle scattering. However in the $p_T$ regime covered by the experimental measurements hadronization could play an important role; coalescence with quarks of the medium, displaying a sizable elliptic flow, could contribute to enhance the heavy-flavor $v_2$~\cite{rapp}.
\section{Conclusions}
We studied the dynamics of $c$ and $b$ quarks in the medium produced in heavy-ion collisions. In our approach, based on the relativistic Langevin equation, the quarks interact with the medium through multiple uncorrelated random momentum-exchanges, which would asymptotically drive them to thermal equilibrium. The heavy-quark transport coefficients were evaluated within a weak-coupling framework. Our procedure leads to results in reasonable agreement with the available experimental data, which extend up to quite high $p_T$. This could even suggest to reconsider the role of collisional energy-loss in the suppression of light-hadron spectra. 

\end{document}